\documentclass[final,5p,times,twocolumn]{elsarticle}

\usepackage{amssymb}
\usepackage{amsmath}
\usepackage{siunitx}
\usepackage{gensymb}
\usepackage{textcomp}
\usepackage{hyperref}

\usepackage{xcolor}
\usepackage{ulem}
\usepackage{soul}

\usepackage{tikz}
\usetikzlibrary{arrows.meta, positioning}
\usepackage{algorithm}
\usepackage[noend]{algpseudocode}

\newcommand{\jt}[1]{#1}
\newcommand{\rev}[1]{#1}

\journal{Computer Physics Communications}

\begin{document}

\begin{frontmatter}



\title{Data-driven reduced modeling of streamer discharges in air}


\author[label1]{Jannis Teunissen}
\author[label1,label2]{Alejandro Malagón-Romero}

\affiliation[label1]{organization={Centrum Wiskunde \& Informatica (CWI)}, 
            addressline={Science Park 123},
            city={Amsterdam},
            postcode={1098 XG},
            country={The Netherlands}}
\affiliation[label2]{organization={Instituto de Astrof{\'\i}sica de Andaluc{\'\i}a (IAA), CSIC},
            addressline={Glorieta de la Astronomía s/n},
            city={Granada},
            postcode={18008},
            country={Spain}}

\begin{abstract}
  We present a computational framework for simulating filamentary electric discharges, in which channels are represented as conducting cylindrical segments.
  The framework requires a model that predicts the position, radius, and line conductivity of channels at a next time step.
  Using this information, the electric conductivity on a numerical mesh is updated, and the new electric potential is computed by solving a variable-coefficient Poisson equation.
  A parallel field solver with support for adaptive mesh refinement is used, and the framework provides a Python interface for easy experimentation.
  We demonstrate how the framework can be used to simulate positive streamer discharges in air.
  First, a dataset of 1000 axisymmetric positive streamer simulations is generated, in which the applied voltage and the electrode geometry are varied.
  Fit expressions for the streamer radius, velocity, and line conductivity are derived from this dataset, taking as input the size of the high-field region ahead of the streamers.
  We then construct a reduced model for positive streamers in air, which includes a stochastic branching model.
  The reduced model compares well with the axisymmetric simulations from the dataset, while allowing spatial and temporal step sizes that are several orders of magnitude larger.
  3D simulations with the reduced model resemble experimentally observed discharge morphologies.
  The model runs efficiently, with 3D simulations with 20+ streamers taking 4--8 minutes on a desktop computer.
\end{abstract}




\end{frontmatter}

\section{Introduction}
\label{sec:introduction}

Streamer discharges are the precursors of lightning leaders and sparks, they occur above thunderstorms as sprites, and they are used in diverse technological applications~\cite{Nijdam_2020}.
Streamers develop non-linearly at velocities of $10^5$ to $10^7$ m/s, forming conducting channels with strong electric field enhancement at their tips.
The enhanced field causes the conductive channels to grow due to rapid electron impact ionization.
Numerical simulations have become a valuable tool to understand and predict streamer discharges.
Simulations are usually performed with a fluid model, see e.g.\ \cite{Bourdon_2010,Becker_2013b,Luque_2012,Komuro_2018,Marskar_2019a,Marskar_2020,Ono_2020}, although particle simulations are sometimes also used, see e.g.\ \cite{Wang_2024,Jiang_2017}.
The cost of such `microscopic' simulations is typically quite high, because a high temporal and spatial resolution is required to resolve the electron dynamics.

The two main processes that drive a streamer discharge are electron impact ionization and electron drift.
\jt{Accurately describing} the spatial growth of the electron density in a fluid or particle model requires a grid spacing
\begin{equation}
  \label{eq:dx-alpha}
  \Delta x \lesssim 1/\jt{|}\bar{\alpha}\jt{|},
\end{equation}
where $\bar{\alpha} = \alpha - \eta$ is the Townsend ionization coefficient minus the attachment coefficient.
Accurately resolving the temporal growth of the electron density generally requires a time step
\begin{equation}
  \label{eq:dt-alpha-vd}
  \Delta t \lesssim 1/|\bar{\alpha} \, \mathbf{v}_d|,
\end{equation}
where $\mathbf{v}_d = -\mu_e \mathbf{E}$ is the electron drift velocity, $\mu_e$ the electron mobility and $\mathbf{E}$ the electric field.
The right-hand side is the inverse of the effective electron impact ionization rate, but it can also be thought of as a resulting from a CFL-like condition $\Delta t \lesssim \Delta x/v_d$, with $\Delta x$ given by equation~\eqref{eq:dx-alpha}.
The above time step restriction also applies to models that use implicit time integration if an accurate solution is required, as discussed in~\cite{Bagheri_2018}.

\begin{figure}
  \centering
  \includegraphics[width=\linewidth]{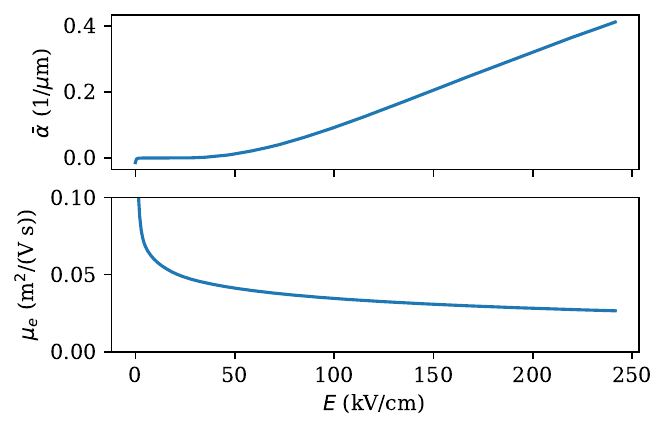}
  \caption{The effective ionization coefficient $\bar{\alpha}$ and electron mobility $\mu_e$ for electrons in air (80\% $\mathrm{N}_2$, 20\% $\mathrm{O}_2$) at 1\,bar and 300\,K, computed from Phelps's cross sections~\cite{Phelps_database,Pitchford_1982}.}
  \label{fig:alpha-mu}
\end{figure}

Figure~\ref{fig:alpha-mu} shows data for $\bar{\alpha}$ and $\mu_e$ for air at 1\,bar and 300\,K.
A typical streamer in air has a maximum electric field $E_\mathrm{max}$ at its tip between 100\,kV/cm and \jt{250}\,kV/cm.
In this range, $\bar{\alpha}$ is approximately proportional to the electric field strength $E$, and so is the drift velocity, since $\mu_e$ is approximately constant.
This means that $\Delta x \propto (E_\mathrm{max})^{-1}$ and $\Delta t \propto (E_\mathrm{max})^{-2}$, so that the cost of a uniform-grid 3D simulation will approximately scale as $(\Delta x)^{-3} \, (\Delta t)^{-1} \sim (E_\mathrm{max})^5$.
Figure~\ref{fig:alpha-mu} shows that $\Delta x$ will range from about $10 \, \mu\textrm{m}$ to $\jt{2.5} \, \mu\textrm{m}$ for fields between 100\,kV/cm and \jt{250}\,kV/cm, whereas streamer discharges in air can measure centimeters or more.
Most 3D simulations are therefore performed with adaptive mesh refinement (AMR), see e.g.\ \cite{Montijn_2006,Pancheshnyi_2008,Kolobov_2012,Teunissen_2017,Marskar_2019a}.
However, even then there are many cases of interest that are currently too expensive to simulate, for example because the discharge contains too many streamer channels, or because the evolution on large spatial or temporal scales is of interest.

Several authors have developed reduced models in which the smallest spatial and temporal scales for electrons do not have to be resolved.
In the dielectric breakdown model (DBM)~\cite{Niemeyer_1984}, the domain is separated into conducting and non-conducting cells.
The conducting region expands stochastically, one cell at a time, with the probability depending on the local electric field strength.
In~\cite{Pasko_2000}, this type of model was modified for the simulation of sprite discharges, and in~\cite{Noskov_2001}, the model was adapted for simulating discharge treeing in solid dielectrics.
A somewhat related model was developed in~\cite{Akyuz_2003}, in which streamers were modeled as segments of perfectly conducting cylinders, and a finite-element solver was used to compute the electric field configuration of the discharge.
To improve some shortcomings of the above models, a `tree model' was presented in~\cite{Luque_2014} and further refined in~\cite{Luque_2017a}.
In these tree models, the channels are approximated by a series of point or ring charges, with the radius, conductivity and charge density evolving along the channel.
The electric field and the electric current are computed self-consistently by including the pair-wise interactions between all point or ring charges.
Although more physics is included than in the DBM models, these tree models still require some ad-hoc parameters, such as a prescription for the streamer radius.

In this paper, we make several contributions to the development of reduced models for electric discharges.
In section~\ref{sec:framework-description} we present a framework for simulating the growth of imperfectly conducting cylinders in a self-consistent electric field.
As input, a growth model for the cylinders and their conductivity is required.
The conductivity of the cylinders is mapped to a numerical mesh, and by solving a modified Poisson equation on this mesh the electric potential at the next time step can be found.
The field solver includes adaptive mesh refinement and support for electrodes using the \texttt{afivo}~\cite{Teunissen_2018} library, and the resulting code is converted to a Python module.
The growth model for the cylinders can be fully written in Python, allowing for easy experimentation with different types of data-driven or physics-based models.

Next, we describe a dataset of 1000 axisymmetric positive streamer simulations in section~\ref{sec:simulation-data-set}.
The simulations are performed in air at 1\,bar and 300\,K, while the applied voltage and the electrode geometry are varied.
From the dataset, we fit a simple model for the streamer radius, velocity and conductivity, taking as input a length scale of the high-field region around the streamer head.
Based on these expressions, we construct a reduced model for streamer growth in section~\ref{sec:model-description}, using the framework described in section~\ref{sec:framework-description}.
With this model, the grid spacing and time step can be much larger than the restrictions given by  equations~\eqref{eq:dx-alpha} and~\eqref{eq:dt-alpha-vd}.
Finally, we compare the model's prediction to the original fluid simulation in the dataset, and we show examples of 3D simulations in section~\ref{sec:results}.

\section{Framework for growing conducting cylinders in an electric field}
\label{sec:framework-description}

Streamers and lightning leaders are examples of filamentary electric discharges, typically consisting of many conductive channels that rapidly grow and sometimes branch.
Below, we will describe a basic framework to simulate such phenomena.
\rev{One assumption underlying our approach is that channels have a highly similar shape with a rounded head, as has been observed in many simulations and experiments.
  This allows us to describe each channel as a collection of conducting cylindrical segments with a semi-spherical cap at the end.
  It is assumed that channels propagate parallel to the electric field ahead of them, and the radial electron density profile inside each channel is assumed to be fixed.}

\rev{Another important assumption is that much of the complex dynamics of a discharge results from electrostatic interactions.
  The framework was designed to capture these electrostatic interactions rather well.
  The streamer conductivity is therefore represented on a numerical mesh that is fine enough to resolve the streamer radii and the electric field profiles around streamer heads.
}
  This mesh is a structured grid with quadtree/octree type grid refinement, and its spacing can be hundreds of $\mu$m for streamer discharges in atmospheric air, as we will show in section~\ref{sec:dependence-time-step}.

It is assumed that some model is available to predict the position $\mathbf{r}$, radius $R$ and line conductivity $\sigma_\mathrm{line}$ of each channel at the next time step.
This information can be used to update the conductivity in the domain, by adding cylindrical segments representing the extension of each channel, as described in section~\ref{sec:mapp-change-cond}.
Afterwards, the electric potential at the new time step can be computed on the numerical mesh as described in section~\ref{sec:evolv-cond-potent}.
\jt{As will be discussed in section~\ref{sec:dependence-time-step}, time steps in the reduced model can be two to three orders of magnitude larger than in conventional streamer simulations.}

\subsection{Conducting cylinders}
\label{sec:mapp-change-cond}

\begin{figure}
  \centering
  \includegraphics[width=\linewidth]{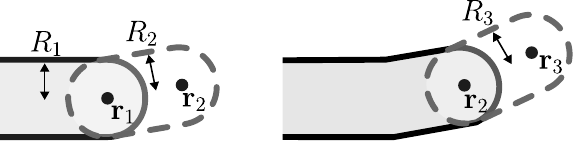}
  \caption{Illustration of how cylindrical segments with a semi-spherical cap can be added together to \rev{extend a channel from $\mathbf{r}_1$ to $\mathbf{r}_2$ (left) and afterwards from $\mathbf{r}_2$ to $\mathbf{r}_3$ (right)}. We assume a constant radius $R$ along an individual segment. Note that the respective $\mathbf{r}$ coordinates are defined at the `center' of the segment's caps.
    \jt{The figure is not to scale: in actual simulations, the length of a newly added segment is comparable to the streamer radius or smaller, see section~\ref{sec:updat-stre-posit}.}
    }
  \label{fig:cylindrical-segment}
\end{figure}

The conducting channels will be represented by cylindrical segments with a semi-spherical cap at the end.
It is assumed that some model is available to predict the following properties:
\begin{itemize}
  \item The next radius $R(t+\Delta t)$.
  \item The next position $\mathbf{r}(t+\Delta t)$, located at the center of the semi-spherical cap, as illustrated in figure~\ref{fig:cylindrical-segment}.
  \item The next line conductivity $\sigma_\mathrm{line}(t+\Delta t)$, where the line conductivity is defined as the integral over the conductivity over the channel's cross-section.
\end{itemize}
Furthermore, we assume there is a function $f_r(d_r/R)$ that describes the radial conductivity profile, where $d_r$ is the distance from the cylinder's axis.
With the above information, a cylindrical segment can be added between $\mathbf{r}(t)$ and $\mathbf{r}(t+\Delta t)$, see figure~\ref{fig:cylindrical-segment}.
To update the conductivity on the mesh, the following conditions are checked for every grid cell:
\begin{itemize}
  \item $d < R(t+\Delta t)$, where $d$ is the distance from the cell center to the nearest point of the line segment between $\mathbf{r}(t)$ and $\mathbf{r}(t+\Delta t)$. This condition is fulfilled inside a cylindrical segment of radius $R(t+\Delta t)$ with semi-spherical caps.
  \item $d_0 > R(t)$, where $d_0$ is the distance \rev{to} the previous position $\mathbf{r}(t)$. This excludes the cap of the previous segment.
  Furthermore, the cell should be closer to $\mathbf{r}(t+\Delta t)$ than it is to $\mathbf{r}(t) - [\mathbf{r}(t + \Delta t) - \mathbf{r}(t)]$, which prevents adding conductivity \jt{behind} the previous cap when the radius increases.
\end{itemize}
When these criteria a met, the change in conductivity in a grid cell is given by
\begin{equation}
  \label{eq:sigma-update}
  \sigma_\mathrm{new} = \sigma_\mathrm{old} + \sigma_\mathrm{line}^* \, \frac{f_r(d_r/R)}{\pi R^2},
\end{equation}
where $\sigma_\mathrm{line}^*$ is obtained by linearly interpolating between $\sigma_\mathrm{line}(t)$ at $\mathbf{r}(t)$ and $\sigma_\mathrm{line}(t+\Delta t)$ at $\mathbf{r}(t+\Delta t)$.
Note that the radial profile $f_r(x)$ should be normalized so that
\begin{equation}
  \label{eq:radial-integral}
  \int_0^\infty 2\pi \, r \, \frac{f_r(r/R)}{\pi R^2} \, dr = 1.
\end{equation}
As discussed in \ref{sec:comp-radius}, we \rev{here use a truncated parabola~\cite{Luque_2014,Luque_2016} for the radial profile}
\begin{equation}
  \label{eq:fr-function}
  f_r(x) = \max\left[0, 2 \, (1 - x^2)\right].
\end{equation}
\jt{We remark that there are several definitions of the streamer radius $R$, e.g., optical and electrodynamic.
In equations \eqref{eq:sigma-update}--\eqref{eq:fr-function} $R$ should be the effective radius of the conductivity profile.
Later in this paper we will refer to this radius as $R_\sigma$.}

\subsection{Evolving conductivity and potential}
\label{sec:evolv-cond-potent}

The procedure described in section~\ref{sec:mapp-change-cond} can be used to obtain the conductivity at the next time step $\sigma(\mathbf{r}, t+\Delta t)$.
Below, we discuss how the new electric potential $\phi(t + \Delta t)$ can be computed using $\sigma(t+\Delta t)$, where have dropped the dependence on $\mathbf{r}$ for brevity.
Recall that the electric potential $\phi$ is the solution of
\begin{equation}
  \label{eq:laplace-eq}
  \nabla^2 \phi = -\rho/\varepsilon_0,
\end{equation}
where $\rho$ is the charge density.
From the conservation of charge, it follows that
\begin{equation}
  \label{eq:dt-rho}
  \partial_t \rho = - \nabla \cdot \mathbf{J} \approx - \nabla \cdot (\sigma \mathbf{E}) = \nabla \cdot (\sigma \nabla \phi),
\end{equation}
where we approximated the current density as $\mathbf{J} \approx \sigma \mathbf{E}$, ignoring the effects of diffusion, and used $\mathbf{E} = - \nabla \phi$.
Integrating equation~\eqref{eq:dt-rho} over time gives
\begin{equation}
  \rho(t+\Delta t) = \rho(t) + \int_{t}^{t+\Delta t} \nabla \cdot [\sigma(t') \nabla \phi(t')] \, \mathrm{d}t',
\end{equation}
so that
\begin{equation}
  \label{eq:phi-update-1}
  \nabla^2 \phi(t+\Delta t) = -\frac{\rho(t)}{\varepsilon_0} -\frac{1}{\varepsilon_0}
  \int_{t}^{t+\Delta t} \nabla \cdot [\sigma(t') \nabla \phi(t')] \, \mathrm{d}t'.
\end{equation}
We use a first-order accurate backward-Euler scheme to solve this equation, by approximating $\phi(t') = \phi(t+\Delta t)$ and $\sigma(t') = \sigma(t+\Delta t)$ inside the integral, which results in
\begin{align}
  \label{eq:phi-update-2}
  \nabla \cdot [\epsilon_\sigma \, \nabla \phi(t+\Delta t)] &= -\rho(t)/\varepsilon_0,\\
  \epsilon_\sigma &= 1 + \Delta t \, \sigma(t+\Delta t)/\varepsilon_0.
\end{align}
This is the same variable-coefficient elliptic PDE as is solved when using a so-called semi-implicit scheme in discharge simulations~\cite{Ventzek_1994,Hagelaar_2000a}.

We solve equation~\eqref{eq:phi-update-2} using the geometric multigrid methods implemented in the \texttt{afivo} library~\cite{Teunissen_2018,Teunissen_2023}.
The \texttt{afivo} library supports OpenMP parallelization, grid refinement (of the quadtree/octree type), and it is possible to include curved electrodes using a level-set function~\cite{Teunissen_2023}.
Equation~\eqref{eq:phi-update-2} is solved using FMG (full multigrid) cycles, until the residual is smaller than $10^{-5} \, \max(|\rho(t)|/\varepsilon_0)$.
Due to the variable coefficient, this typically takes about 4-6 FMG cycles.

\jt{After $\phi(t+\Delta t)$ has been obtained, we use equation~\eqref{eq:laplace-eq} to compute $\rho(t+\Delta t)$.
  The degree to which charge is conserved, except for currents through the domain boundaries, depends on the accuracy with which equation~\eqref{eq:phi-update-2} is solved.
  If the Poisson equation is solved up to machine precision, charge will be conserved approximately up to machine precision.
}

\subsection{Implementation as a Python module}
\label{sec:implementation}

To make it easier to experiment with different types of models, we have separated the implementation into a Fortran and a Python part.
The computationally expensive parts are all implemented in Fortran, for example the solution of equation~\eqref{eq:phi-update-2}, and the updating of the electric conductivity on the mesh.
With \texttt{F2PY} (Fortran to Python interface generator), this Fortran code was converted to a Python module.
The most important methods available through this module are:
\begin{itemize}
  \item A method to update the conductivity in the domain, given the new and previous channel positions $\mathbf{r}_i$, radii $R_i$ and line conductivities $\sigma_{\mathrm{line}, i}$.
  \item A method to solve the Poisson equation given by ~\eqref{eq:phi-update-2}.
  \item Functions to initialize the computational domain (size, minimum grid spacing, refinement criteria), the applied voltage, and the electrode geometry.
  \item A method to store a field-dependent effective ionization rate, which is used to update the conductivity inside channels according to equation~\eqref{eq:sigma-deriv}.
  \item Methods to get the electric field vector at a location, to get the location of the maximum field, and to interpolate a variable (electric potential, electric field strength, conductivity) along a line on the mesh.
  \item A method to update the adaptive mesh refinement.
  \item A method to write the mesh variables to a Silo file, that can be visualized using e.g.\ Visit~\cite{Childs_2012}.
\end{itemize}

The framework can be used in both 2D and 3D.
To avoid duplicating code for 2D and 3D, a header file with several macros is included in the Fortran source code (provided by the \texttt{afivo} library~\cite{Teunissen_2018}), so that the number of dimensions can be specified at compile time.
By default, no mesh refinement is performed in 2D, since 2D simulations are computationally not expensive.

\section{Simulation dataset}
\label{sec:simulation-data-set}

\rev{The framework described in section~\ref{sec:framework-description} requires a method that predicts the next radius, velocity and line conductivity of each channel.
  For this purpose,} we have constructed a dataset of axisymmetric simulations of positive streamers in air by varying the applied voltage and electrode geometry, \rev{from which we derive simple fit formulas for streamer parameters.}

\subsection{Fluid model and simulation conditions}
\label{sec:fluid-model-simul}

The simulations are performed with the classical drift-diffusion-reaction fluid model using the open-source \texttt{afivo-streamer} code \cite{Teunissen_2018,Teunissen_2017}.
In this model, the electron density $n_e$ evolves in time as
\begin{equation}
    \label{eq:ddt_e}
	\partial_t n_e = \nabla \cdot (\mu_e \mathbf{E} n_e + D_e \nabla n_e) + S_e + S_{\mathrm{ph}},
\end{equation}
where $\mu_e$ is the electron mobility coefficient, $D_e$ is the diffusion coefficient, $\mathbf{E}$ is the electric field, $S_e$ is a source (and sink) term of free electrons due to reactions, and $S_{\mathrm{ph}}$ is the photo-ionization source term.
The motion of ions not taken into account, so their densities only change in time due to source terms.
The simulations are performed in artificial air, consisting of 80\% $\mathrm{N}_2$ and 20\% $\mathrm{O}_2$ at 1 bar and 300 K.
We use the same reactions and transport data as e.g.\ \cite{Li_2021}, computed using BOLSIG-~\cite{Hagelaar_2005} using Phelps' cross-section data for $\mathrm{N}_2$ and $\mathrm{O}_2$~\cite{Phelps_database,Pitchford_1982}.
The photoionization source term $S_{\mathrm{ph}}$ is computed using the Zheleznyak model of \cite{Zheleznyak_1982} and the Helmholtz approximation, see e.g.\ \cite{Bagheri_2018}.

\begin{figure}
  \centering
  \includegraphics[width=8cm]{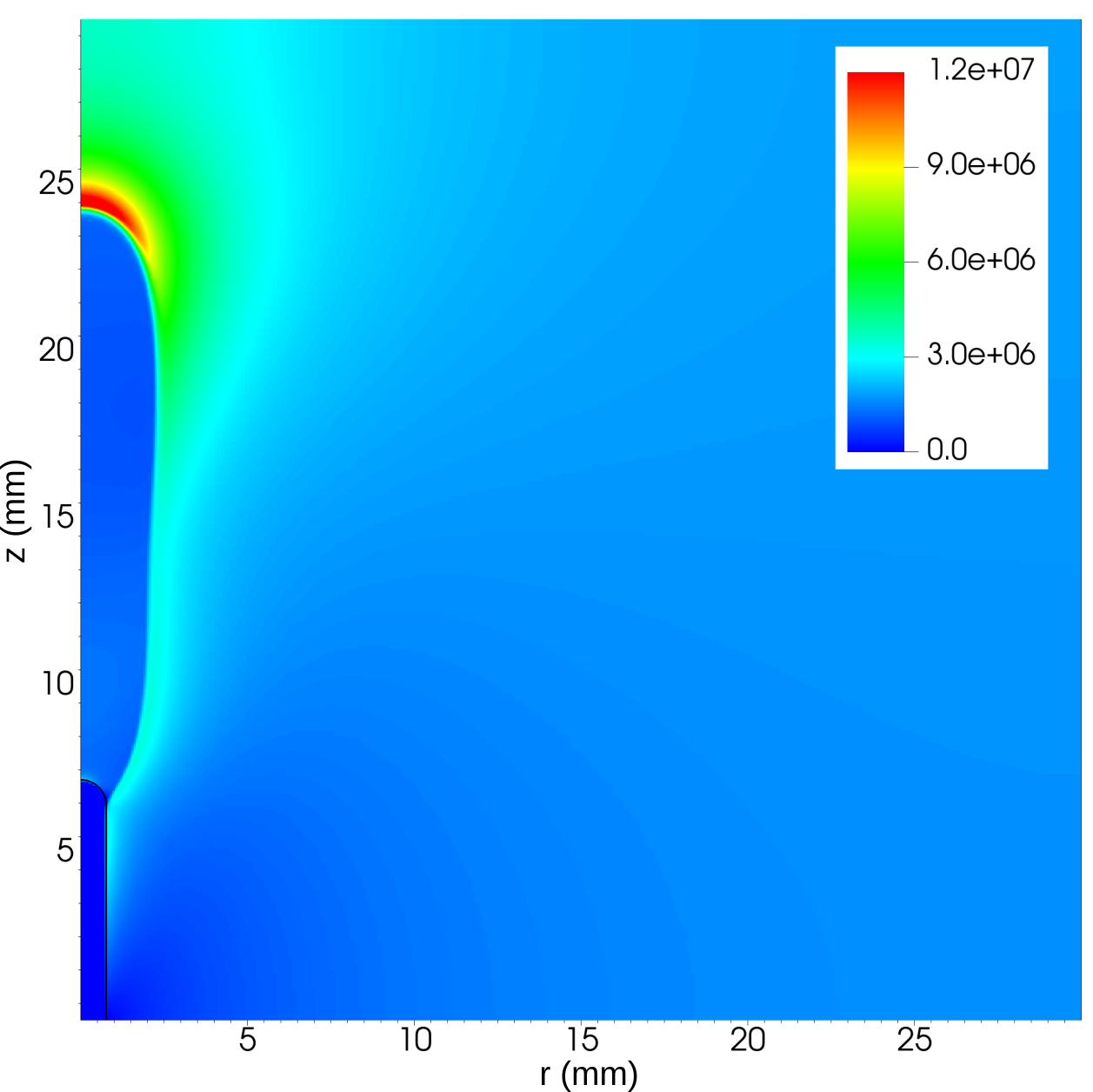}
  \caption{Example of the electric field distribution in one of the axisymmetric streamer simulations.
    The boundary of the needle electrode at the bottom left is marked with a line.}
  \label{fig:comp-domain}
\end{figure}

The axisymmetric computational domain is illustrated in figure~\ref{fig:comp-domain}.
It measures $30 \, \textrm{mm}$ in the $r$ and $z$ direction, and it has a plate-plate geometry.
A rod-shape needle electrode placed at the bottom plate provides initial field enhancement so that a positive streamer can start.
This needle is shaped as a cylinder with a semi-spherical cap, and its length and radius are varied as described in section~\ref{sec:description-data-set}.
In order for discharges to start, a small background ionization density $n_0 = 10^{11} \, \textrm{m}^{-3}$ of electrons and $\mathrm{N}_2^+$ ions is included.

The electric field $\mathbf{E}$ is calculated as $\mathbf{E} = -\nabla \phi$, where $\phi$ is the electrostatic potential, obtained by solving the Poisson equation with a multigrid method~\cite{Teunissen_2018,Teunissen_2023}.
The electric potential is fixed on the bottom and top of the domain, with the needle electrode at the same voltage as the bottom plate.
Homogeneous Neumann boundary conditions are used on the radial boundary.

At the needle electrode, electrons are absorbed but not emitted.
On the domain boundaries, homogeneous Neumann boundary conditions are used for the electron density, but simulations stop before this becomes relevant.
All simulations are performed with adaptive mesh refinement (AMR), using the refinement criterion $\alpha(E) \Delta x \leq 1$, where $\alpha(E)$ is the Townsend ionization coefficient.
\jt{The mesh refinement provided by the \texttt{afivo} library \cite{Teunissen_2018} is of the quadtree/octree type, with a refinement ratio of two between levels.
Time integration is performed explicitly with a second-order accurate Runge-Kutta scheme, so the time step $\Delta t$ is limited by a CFL (Courant–Friedrichs–Lewy) condition, see~\cite{Teunissen_2017} for details.
The finest grid spacing required for the dataset described below was $\Delta x = 1.8\,\mu \mathrm{m}$, although for most cases $\Delta x = 3.7\,\mu \mathrm{m}$ was sufficient.
For $\Delta x = 3.7\,\mu \mathrm{m}$, time steps were typically on the order of $\Delta t \sim 2 \, \mathrm{ps}$.
}

\subsection{Description of dataset}
\label{sec:description-data-set}

\begin{figure*}
  \centering
  \includegraphics[width=1.0\linewidth]{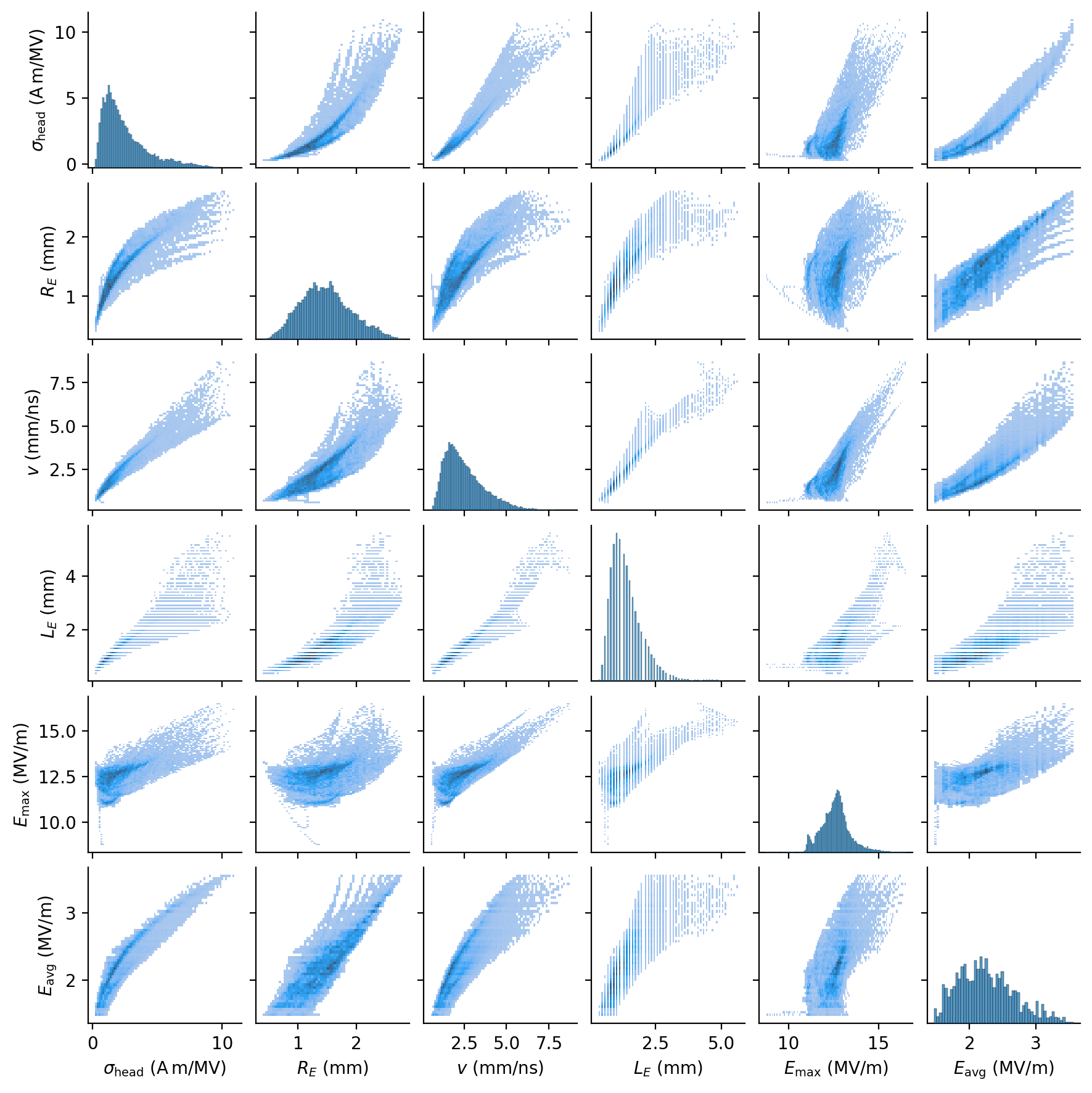}
  \caption{Pairwise relationships between parameters extracted from 1000 axisymmetric positive streamer simulations in air at 1\,bar and 300\,K. On the diagonal the distributions of the parameters are shown.}
  \label{fig:pairplot}
\end{figure*}

We generate a dataset of 1000 axisymmetric streamer simulations in which the following parameters are varied:
\begin{itemize}
  \item The radius $R_\mathrm{rod}$ of the rod electrode \rev{is} drawn from a uniform distribution between $R_\mathrm{min}=0.5 \, \textrm{mm}$ and $R_\mathrm{max}=1.5\,\textrm{mm}$
  \item The length of the rod electrode \rev{is} scaled linearly with its radius as
  \begin{equation*}
    L_\mathrm{rod} = L_\mathrm{min} + (L_\mathrm{max} - L_\mathrm{min}) \frac{R_\mathrm{rod} - R_\mathrm{min}}{R_\mathrm{max} - R_\mathrm{min}},
  \end{equation*}
  using $L_\mathrm{max} = 12 \, \mathrm{mm}$ and $L_\mathrm{min} = 4.5 \, \mathrm{mm}$.
  \item The applied voltage $V_0$ \rev{is} drawn from a uniform distribution between $V_\mathrm{min} = 36 \, \mathrm{kV}$ and $V_\mathrm{max} = 60 \, \mathrm{kV}$.
\end{itemize}
With these applied voltages, the background electric fields $E_\mathrm{bg}$ between the two parallel plates range from $12 \, \mathrm{kV/cm}$ to $20 \, \mathrm{kV/cm}$.
Each simulation \rev{is} stopped when the distance to the opposite electrode \rev{is} $5 \, \textrm{mm}$.
Streamer velocities vary significantly with the above parameters.
To obtain about 30 output files per simulation, we empirically scale the time step for writing output $\Delta t_\mathrm{output}$ with $E_\mathrm{avg}^{-2}$, where $E_\mathrm{avg}$ is the average electric field between the tip of the rod electrode and the opposite plate electrode.
The dataset \rev{consists of} approximately $3 \times 10^4$ output files, which contain the species densities, electric field and electric potential in the full simulation domain.
To allow for easier processing, these files (with AMR) are converted to a uniform resolution of $256^2$.
We extract several quantities from the simulation data:
\begin{itemize}
  \item The maximum electric field at the streamer head $E_\mathrm{max}$.
  \item The $z$-coordinate where $E_\mathrm{max}$ occurs, called $z_\mathrm{head}$.
  \item The streamer velocity $v$, defined as
  \begin{equation*}
    \rev{v = \left[z_\mathrm{head}(t) - z_\mathrm{head}(t-\Delta t)\right]/\Delta t,}
  \end{equation*}
  where $\Delta t$ is the time step between writing output.
  \item The streamer electrodynamic radius $R_E$, defined as the radial coordinate at which the radial electric field $E_r$ has a maximum. \jt{Note that $R_E$ is determined over the entire streamer head, not restricted to $z = z_\mathrm{head}$.}
  \item The effective radius $R_\sigma$ of the streamer's radial conductivity profile. As discussed in~\ref{sec:comp-radius}, we approximate this radius as
  \begin{equation}
    \label{eq:R-sigma}
    R_\sigma = 1.2 \, R_E.
  \end{equation}
  \item The line conductivity due to electrons
  \begin{equation}
    \label{eq:line-sigma}
    \sigma(z) = 2 \pi e \int_0^\infty r \, \mu_e(r, z) \, n_e(r, z) \, dr,
  \end{equation}
  where $e$ is the elementary charge. We define line conductivity at the head as
  \begin{equation}
    \label{eq:sigma-head}
    \sigma_\mathrm{h} = \sigma(z_\mathrm{head} - R_E),
  \end{equation}
  \jt{for a streamer propagating in the $+z$ direction, since we observed that the maximum line conductivity typically occurred about one radius behind $z_\mathrm{head}$.}
  \item The size $L_E$ of the high-field region ahead of a streamer, defined as the distance between $z_\mathrm{head}$ and the $z$-coordinate where the electric field drops below a threshold of $E_\mathrm{threshold} = 50 \, \mathrm{kV/cm}$.
\end{itemize}
Figure~\ref{fig:pairplot} shows the distributions and relations between the above-listed parameters in the dataset.
Note that $\sigma_\mathrm{h}$ and $L_E$ were determined from the $256^2$ uniform data, whereas the other parameters were extracted from the original simulation files.
Because the measurements of $R_E$ are somewhat noisy, we applied a Savitzky-Golay filter of width 5 and order 2 to the data from each run.
Furthermore, for each run, the first two outputs were excluded from the dataset.

$L_E$ is a relevant length scale that can also be determined on coarse grids, as discussed in section~\ref{sec:determining-l_e}.
The threshold used to determine $L_E$ is somewhat arbitrary, and we have experimented with several values ranging between $45 \, \textrm{kV/cm}$ and $60 \, \textrm{kV/cm}$.
Higher thresholds improved the ability to predict streamer properties from $L_E$ (see section \ref{sec:simple-model}), but since $L_E$ was then smaller, they also resulted in more noise when $L_E$ was determined on a numerical grid.

\subsection{A simple model for $v$, $R$ and $\sigma_\mathrm{h}$}
\label{sec:simple-model}

From the dataset, we now construct a simple model to predict the electrodynamic radius $R_E$, the velocity magnitude $v$ and the line conductivity at the streamer head $\sigma_\mathrm{h}$.
There are many ways to fit the curves shown in figure~\ref{fig:pairplot}, but we here opt for the following approximations:
\begin{align}
  \label{eq:model-sigma}
  \sigma_\mathrm{h} &= \begin{cases}
    \num{1e-8} + 1.40 \, L_E^2 & L_E < 10^{-3}\\
    \num{-1.41e-06} + \num{2.80e-3} \, L_E & L_E \geq 10^{-3}
  \end{cases}
  \\
  \label{eq:model-R}
  R_E &= \begin{cases}
    \num{2.90e-05} + {1.30} \, L_E & L_E < 10^{-3}\\
    \num{6.31e-4} + 0.627 \, L_E & L_E \geq 10^{-3}
  \end{cases}
  \\
  v &= \num{1.78e+09} \, L_E,\label{eq:model-v}
\end{align}
where the quantities are all made dimensionless using the following units: m for $R_E$ and $L_E$, A\,m/V for $\sigma_\mathrm{h}$, V/m for $E_\mathrm{avg}$ and m/s for $v$.
Note that the expression for $\sigma_\mathrm{h}$ consists of a quadratic function for $L_E < 1 \, \mathrm{mm}$, while for $L_E \geq 1 \, \mathrm{mm}$ the tangent line at $L_E = 1 \, \mathrm{mm}$ is used.
For $R_E$ a broken line is used, with the break also at $L_E = 1 \, \mathrm{mm}$, and for $v$ the fit is a single line.
The reason for using `broken curves in these fits is that we wanted to approximate both the behavior at small $L_E$ and at large $L_E$ reasonably well, while ensuring non-negativity of $\sigma_\mathrm{h}$, $R_E$ and $v$ for $L_E \geq 0$.
\rev{Furthermore, note that all expressions are linear in $L_E$ for $L_E > 1 \, \mathrm{mm}$.}

\begin{figure}
  \centering
  \includegraphics[width=0.9\linewidth]{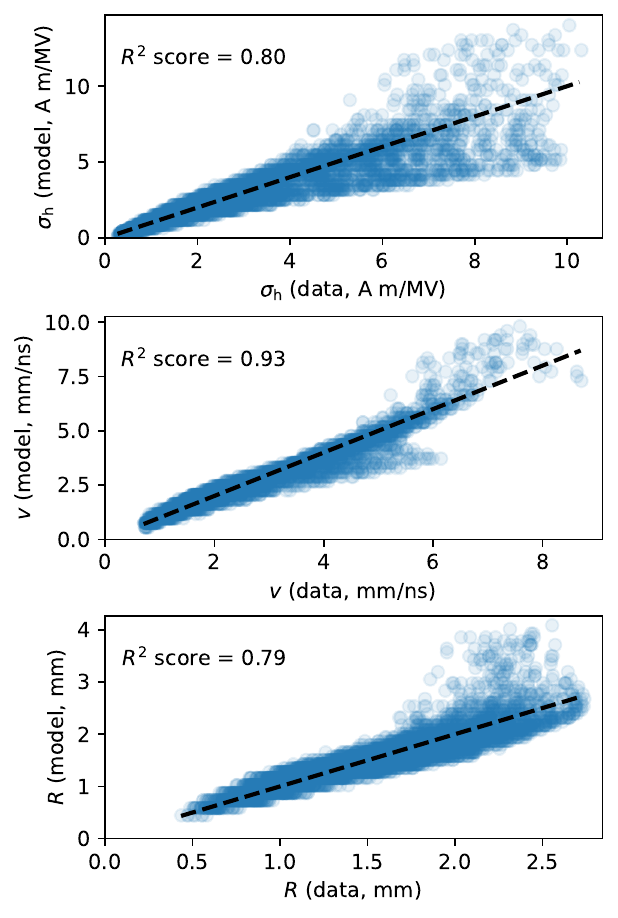}
  \caption{A comparison of the simple model given by equations~\eqref{eq:model-sigma}--\eqref{eq:model-v} against the simulation dataset.
    The model predictions are shown on the vertical axes, and the data on the horizontal axes.
    The dashed lines indicate were predictions and data are equal.}
  \label{fig:r-v-sigma-model}
\end{figure}

In figure~\ref{fig:r-v-sigma-model}, equations~\eqref{eq:model-sigma}--\eqref{eq:model-v} are compared against the dataset.
The $R^2$ scores (coefficients of determination) are also given in the figure, with $1.0$ being the highest possible value if there is no prediction error.
Note that for large values of $L_E$, some predictions of $R_E$ lie a bit further above the data.
These points correspond to the moment the streamer is close to the opposite electrode, which causes $L_E$ to increase faster than the radius.

To fit the above formulas, the data was split in a training set (70\% of the runs) and a test set (30\% of the runs).
However, due to the limited number of parameters in the model, there was essentially no overfitting and thus almost no difference when we compared against the training or the test set.

\subsection{\jt{Comments on dataset}}
\label{sec:dataset-comments}

\jt{The dataset described here has several limitations, even when only considering positive streamers in air.
  Due to the use of an axisymmetric model, the dataset does not contain streamer branching or streamer-interaction phenomena.
  Furthermore, the dataset contains no simulations in low background fields, in which streamers obtain a small radius and possibly stagnate~\cite{Nijdam_2020}, since such simulations can be somewhat challenging to perform~\cite{Pancheshnyi_2004,Niknezhad_2021,Li_2022}.
  Another limitation is that only a single type of electrode geometry was considered, namely two plates with a needle on one side and a fixed gap size.

  With enough computational resources, one could construct a more general dataset based on 3D streamer simulations, including streamer branching, different electrode geometries, and different gap sizes.
  Since negative and positive streamers often form simultaneously, for example in a needle-to-needle geometry, it would make sense to include negative streamers in the dataset \rev{as well}.
  The main challenges in constructing such a dataset are that the simulations are computationally expensive, that a significant amount of data needs to be stored, and that it is harder to extract features from such 3D data.
}

\section{Reduced model for streamer discharges}
\label{sec:model-description}

We now describe a new reduced model for streamer discharges\jt{, which we have named \texttt{cocydimo} for ``Conducting Cylinder Discharge Model''.}
The model makes use of the framework described in section~\ref{sec:framework-description} and the dataset described in section~\ref{sec:simple-model}.
\jt{A schematic overview of the model is shown in algorithm~\ref{alg:pseudocode}, and its main parameters are summarized in table~\ref{tab:variables}.}

In a streamer discharge, the spatial conductivity profile $\sigma(\mathbf{r})$ rapidly changes at streamer heads due to electron impact ionization, and it changes more slowly inside the channels due to e.g.\ electron attachment.
Below, we explain how we predict the conductivity at the next time step $\sigma(\mathbf{r}, t+\Delta t)$ from the previous conductivity $\sigma(\mathbf{r}, t)$, electric potential $\phi(\mathbf{r}, t)$ and the following streamer parameters:
\begin{itemize}
  \item The radius $R_\sigma$ of the streamer's radial conductivity profile.
  \item The position $\mathbf{r}$ at the `center' of the streamer head, so that its conductivity is nonzero within a distance $R_\sigma$ from $\mathbf{r}$.
  \item The propagation direction $\mathbf{\hat{v}}$, where the hat denotes a unit vector.
\end{itemize}
Once $\sigma(\mathbf{r}, t+\Delta t)$ is known, $\phi(\mathbf{r}, t+\Delta t)$ can be computed, as described in section~\ref{sec:evolv-cond-potent}.

\begin{algorithm}
  \caption{Pseudocode for reduced discharge model.}
  \label{alg:pseudocode}
  \begin{algorithmic}[1]
    \State \textbf{Input:} Specify computational domain, electrode geometry, applied voltage, time step, (finest) grid spacing
    \State \textbf{Initialize:} Set streamer start location(s), $t = 0$
    \State

    \While{$t < t_\mathrm{end}$}
    \For{each channel}

    \If{3D}
    \Comment{Sec.~\ref{sec:branching-model}}
    \State Sample $\tau_\mathrm{branch}$
    \If{$\tau_\mathrm{branch} < \Delta t$}
    \State Add new channel
    \State Sample branching angles
    \State Rotate \rev{channel directions}
    \EndIf
    \EndIf

    \State
    \State Determine electric field direction $\mathbf{\hat{E}}$
    \Comment{Sec.~\ref{sec:determ-stre-prop}}

    \State Measure $L_E$ and apply smoothing
    \Comment{Sec.~\ref{sec:determining-l_e}}

    \State Get new $\sigma_\mathrm{h}$ and $v$ from model
    \Comment{Sec.~\ref{sec:simple-model}}

    \State Determine new location $\mathbf{r}$
    \Comment{Sec.~\ref{sec:updat-stre-posit}}
    \EndFor

    \State
    \State Update mesh refinement (in 3D)
    \Comment{Sec.~\ref{sec:comp-against-exper}}
    \State Map conductivity of new segments to mesh
    \Comment{Sec.~\ref{sec:mapp-change-cond}}
    \State Update conductivity in channels
    \Comment{Sec.~\ref{sec:update-channel-conductivity}}
    \State Compute new electric potential and field
    \Comment{Sec.~\ref{sec:evolv-cond-potent}}
    \State $t = t + \Delta t$
    \EndWhile
  \end{algorithmic}
\end{algorithm}

\begin{table*}
  \centering
  \begin{tabular}{|l|l|l|}
    \hline
    \textbf{Variable / parameter} & \textbf{Description} & \textbf{Value / given by} \\ \hline
    $L_E$ & Size of the high-field region ahead of streamer & Eq.~\eqref{eq:L-E-beta}\\
    $E_\mathrm{threshold}$ & Threshold electric field for $L_E$ determination & 50 kV/cm \\
    $\beta$ & Smoothing coefficient for $L_E$ in eq.~\eqref{eq:L-E-beta} & 0.5 \\
    $c_1$ & Coefficient correcting for grid spacing bias in $L_E$ in eq.~\eqref{eq:L-E-corrected} & 0.75 \\
    $R_\sigma(t)$ & Radius of streamer's conductivity profile & Eqs.~\eqref{eq:R-sigma}, \eqref{eq:model-R} \\
    $\sigma_\mathrm{h}(t)$ & Line conductivity at streamer head & Eq.~\eqref{eq:model-sigma} \\
    $\mathbf{r}(t)$ & Position at `center' of streamer head & Eq.~\eqref{eq:position-update} \\
    $\mathbf{v}(t)$ & Streamer velocity & Eqs.~\eqref{eq:model-v}, \eqref{eq:direction-formula}\\
    $c_\mathrm{ahead}$ & Coefficient for sampling field direction in eqs.~\eqref{eq:r-E-definition}, \eqref{eq:direction-formula} & 0.5 \\
    $S(E)$ & Effective ionization rate for updating channel conductivity in eq.~\eqref{eq:sigma-update-channel} & Input data\\
    $c_b$ & Parameter for branching time in eq.~\eqref{eq:branch-tau} & 10 -- 20\\
    $L_b$ & Parameter for branching dependence on radius in eq.~\eqref{eq:branch-tau} & 0.2\,mm -- 0.8\,mm\\
    $\Delta t$ & Time step of the model, see section~\ref{sec:dependence-time-step} & up to 1 ns\\
    $\Delta x$ & Grid spacing for channels, see section~\ref{sec:dependence-time-step} & up to hundreds of $\mu$m\\
    \hline
  \end{tabular}
  \caption{\jt{Summary of key variables and parameters in the reduced model.}}
  \label{tab:variables}
\end{table*}

\subsection{Determining the streamer propagation direction}
\label{sec:determ-stre-prop}

\begin{figure}
  \centering
  \includegraphics[width=5cm]{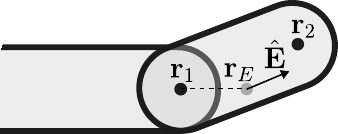}
  \caption{Schematic illustration of the approach \rev{used to determine the propagation direction of a streamer}.
    The electric field unit vector $\mathbf{\hat{E}}$ is sampled at $\mathbf{r}_E$ ahead of the channel.
    \rev{The new conducting segment then points towards $\mathbf{\hat{E}}$.}}
  \label{fig:adding-segment}
\end{figure}

In 3D, we assume that streamers propagate in the direction of the electric field ahead of them.
This field is sampled at a location slightly in front of a streamer, given by
\begin{equation}
  \label{eq:r-E-definition}
  \mathbf{r}_E = \mathbf{r}(t) + (1 + c_\mathrm{ahead}) \, R_\sigma \, \mathbf{\hat{v}}(t),
\end{equation}
where we use $c_\mathrm{ahead} = 0.5$.
There is a balance here: on the one hand, we want the field direction close to the streamer head, but if we get too close to the conductive region, this field direction will be rather noisy on a coarse mesh.
\rev{However, the choice of $c_\mathrm{ahead} = 0.5$ is somewhat arbitrary, and this value could be refined by carefully comparing streamer paths against conventional 3D simulations.
After $\mathbf{r}_E$ has been determined}, the new propagation direction is given by
\begin{equation}
  \label{eq:direction-formula}
  \mathbf{\hat{v}}(t+\Delta t) = \mathbf{\hat{E}}(\mathbf{r}_E),
\end{equation}
as illustrated in figure~\ref{fig:adding-segment}.
In 2D axisymmetric simulations, streamers are simply assumed to propagate in the $z$ direction.

\subsection{Determining $L_E$}
\label{sec:determining-l_e}

To determine $L_E$, we sample the electric field strength along a line from $\mathbf{r}(t)$ extending towards the new propagation direction $\mathbf{\hat{v}}(t+\Delta t)$.
Along this line, $L_E$ is determined as the distance between the location of the maximum field to the location where $E \leq E_\mathrm{threshold}$.
Such measurements of $L_E$ are noisy, with fluctuations on the order of the grid spacing $\Delta x$, which can be quite large in the reduced model.
To reduce these fluctuations, we use exponential smoothing for all but the first measurement:
\begin{equation}
  \label{eq:L-E-beta}
  L_{E, \mathrm{new}} = \beta \, L_{E, \mathrm{new}}^* + (1-\beta) \, L_{E, \mathrm{old}},
\end{equation}
where $L_{E, \mathrm{new}}^*$ is the new measurement and where we use a smoothing factor $\beta$.
\rev{The value of $\beta$ can be specified when running the model; unless stated otherwise we use $\beta = 0.5$.}
In section~\ref{sec:dependence-time-step} we will discuss the dependence of $L_E$ on the grid spacing $\Delta x$ and how to correct for that.

Positive streamers can stagnate~\cite{Pancheshnyi_2004,Niknezhad_2021,Li_2022} when the background field they propagate in becomes too low.
This phenomenon is common in 3D simulations in which some branches are overtaken by others.
We assume streamers have stagnated when $L_E$ drops below a threshold of \rev{$L_{E,\mathrm{min}}$},  and then halt their growth.
\rev{For the simulations presented here we use $L_{E,\mathrm{min}} = 0.1 \, \textrm{mm}$, which corresponds to a minimal radius $R_{E, \mathrm{min}} = 0.15 \, \mathrm{mm}$.
  Smaller radii have been observed in experiments and simulations~\cite{Nijdam_2010, Li_2022}, but only for rather slow `minimal' streamers that are not represented in our dataset.}

\subsection{Updating the streamer position}
\label{sec:updat-stre-posit}

Once $L_E$ is known, equations~\eqref{eq:model-sigma}--\eqref{eq:model-v} are used to obtain new values for $\sigma_\mathrm{h}$, $R_\sigma = 1.2 \, R_E$ and $v$.
We limit the change $\Delta R_\sigma$ during a time step $\Delta t$ to be at most $v \, \Delta t$, which avoids an instantaneous expansion of the radius during streamer inception.
The new streamer position is computed as
\begin{equation}
  \label{eq:position-update}
  \mathbf{r}(t+\Delta t) = \mathbf{r}(t) + \Delta t \, \mathbf{v}(t+\Delta t) - \Delta R_\sigma \mathbf{\hat{v}}(t+\Delta t),
\end{equation}
where $\mathbf{v}(t+\Delta t) = v(t+\Delta t) \, \mathbf{\hat{v}}(t+\Delta t)$.
The change in radius $\Delta R_\sigma$ is subtracted because $\mathbf{r}(t)$ is the position at the center of the streamer head, while we want the conductive region to grow with a given velocity.

\jt{In section \ref{sec:dependence-time-step} we will show that the reduced model can be used with a time step on the order of a nanosecond.
  If we assume a velocity of a few mm/ns and a radius of a few mm based on figure~\ref{fig:pairplot}, the length of a newly added cylindrical segment (about $v \Delta t$) is then comparable to the streamer radius.
}

\subsection{Updating the conductivity in the channels}
\label{sec:update-channel-conductivity}

For simplicity, we here assume that the conductivity in the discharge channels changes according to \begin{equation}
  \label{eq:sigma-deriv}
  \partial_t \sigma(\mathbf{r}, t) = \sigma(\mathbf{r}, t) \, S(E),
\end{equation}
where $S(E) = \bar{\alpha} \, \mu_e E$ is the electric-field dependent effective ionization rate that describes the decay (or growth) of the electron density.
This approximation is justified when the conductivity is dominated by the contribution from electrons, \jt{if the electron mobility in the channel does not vary significantly, }and if the main electron gain and loss processes are electron impact ionization and attachment.
The change in $\sigma$ due to equation \eqref{eq:sigma-deriv} is computed as
\begin{equation}
  \label{eq:sigma-update-channel}
  \jt{\sigma(t+\Delta t) = e^{S \, \Delta t} \sigma(t).}
\end{equation}

Equation~\eqref{eq:sigma-deriv} should not be used \jt{in the high-field region near the streamer head, since the conductivity there is determined by interpolating the line conductivity given by equation~\eqref{eq:model-sigma} to the mesh using equation~\eqref{eq:sigma-update}.} We therefore keep track of the time at which grid cells have become part of any conductive channel, and only update those grid cells which have been inside a channel for a \jt{time $\tau_\mathrm{delay}$ or more.
For the examples presented in this paper, we use $\tau_\mathrm{delay} = 1 \, \mathrm{ns}$.}

\subsection{Branching}
\label{sec:branching-model}

In 3D, streamer discharges typically include multiple channels due to branching.
In $\mathrm{N}_2$-$\mathrm{O}_2$ mixtures, the probability of branching has been shown to depend on the amount of photoionization and the applied voltage~\cite{Nijdam_2020,Briels_2008,Nijdam_2010,Wang_2023,Guo_2024}.
Some observations can be made based on these papers:
First, branching is more likely to occur per unit length when a streamer decelerates and when its radius $R$ decreases.
Second, for thinner streamers the ratio $L_\mathrm{branch}/R$ becomes larger, where $L_\mathrm{branch}$ is the distance between consecutive branches. Below some minimum radius $R$, branching is unlikely.

Like~\cite{Luque_2014}, we here assume that branching can be described as a Poisson process, so that the time $\tau_\mathrm{branch}$ until the next branching event can be sampled from the exponential distribution.
We approximate \jt{the expected value of} $\tau_\mathrm{branch}$ as:
\begin{equation}
  \label{eq:branch-tau}
  \jt{\bar{\tau}}_\mathrm{branch} = c_b \, \frac{R_\sigma}{v} \left(1 + \frac{L_b^2}{R_\sigma^2} \right)
\end{equation}
where $c_b$ and $L_b$ are parameters.
The parameter $L_b$ controls how much $\jt{\bar{\tau}}_\mathrm{branch}$ increases for a streamer with a small radius.
For example, the branching time increases by a factor of two for $R_\sigma = L_b$, and when $R_\sigma = L_b/2$ the increase is a factor five.
In case $R_\sigma \gg L_b$, $c_b$ is approximately the ratio of the distance between branching ($L_\mathrm{branch} = v \, \jt{\bar{\tau}}_\mathrm{branch}$) and the streamer radius.
In the experiments of~\cite{Briels_2008} the ratio $L_\mathrm{branch}/D$ was found to be about $11 \pm 4$, with $D$ being the full width half maximum (FWHM) optical diameter.
Based on the results presented in~\ref{sec:comp-radius} we estimate that \rev{$R_\sigma \approx 0.75 \, \mathrm{FWHM}$}, so that $L_\mathrm{branch}/R_\sigma$ becomes about $15 \pm 5$.

Since the exponential distribution is memoryless, \jt{a new value for $\tau_\mathrm{branch}$ can be sampled from the exponential distribution with rate parameter $\lambda = 1/\bar{\tau}_\mathrm{branch}$} every time step, and if $\tau_\mathrm{branch} \leq \Delta t$, branching occurs.
In a branching event, an axis perpendicular to the parent branch velocity $\mathbf{v}$ is randomly sampled.
An angle $\gamma$ is also sampled, assumed to be uniformly distributed between 0 and 90\degree.
The initial velocities of the two new branches are obtained by rotating $\mathbf{v}$ around the axis over angles $\gamma$ and $\gamma - 90\degree$ respectively.
However, these velocities are only used to determine the electric field directions $\mathbf{\hat{E}}(\mathbf{r}_E)$ ahead of the branches according to equations~\eqref{eq:r-E-definition} and \eqref{eq:direction-formula}, and the branches propagate in the direction of $\mathbf{\hat{E}}(\mathbf{r}_E)$.
The actual branching angle therefore differs from $\gamma$.

\jt{The parameters for the branching model described above were inspired by recent experimental work on streamer branching in N$_2$-O$_2$ mixtures~\cite{Guo_2024}.
  Figure~6 of this paper shows that in their experiments in air the typical angle between two new branches was about 90\degree, and this figure also suggests that the choice of angles $\gamma$ and $\gamma - 90\degree$ is reasonable in air.}

\section{Results}
\label{sec:results}

\subsection{Comparison against axisymmetric simulations}
\label{sec:comp-against-axisymm}

We first compare the reduced model against simulations from the dataset described in section~\ref{sec:description-data-set}.
For the reduced model, the initial condition is zero conductivity in the whole domain.
This still allows a discharge to form since equation~\eqref{eq:model-sigma} depends only on $L_E$, whereas in the fluid simulations some initial electrons are required.
The time step used in the reduced model was set equal to the time between writing output ($\Delta t_\mathrm{output}$) in the fluid simulations, so that it takes about 30 steps to cross the simulation domain as discussed in section~\ref{sec:fluid-model-simul}.
A spatial resolution of $\Delta r = \Delta z = 0.12 \, \mathrm{mm}$ was used, with a uniform mesh of $256^2$ cells.
Adaptive mesh refinement was not used for these cases, since the simulations already run in a few seconds on a uniform grid.
As an initial condition, a streamer with a radius $R_{\sigma,0} = 0.5 \, R_{\sigma}(L_E)$ is placed at the electrode tip, where $R_{\sigma}(L_E)$ is obtained according to equations~\eqref{eq:model-R} and~\eqref{eq:R-sigma}, \rev{while $L_E$ is determined by the electric field profile ahead of the electrode.}

\begin{figure}
  \centering
  \includegraphics[width=\linewidth]{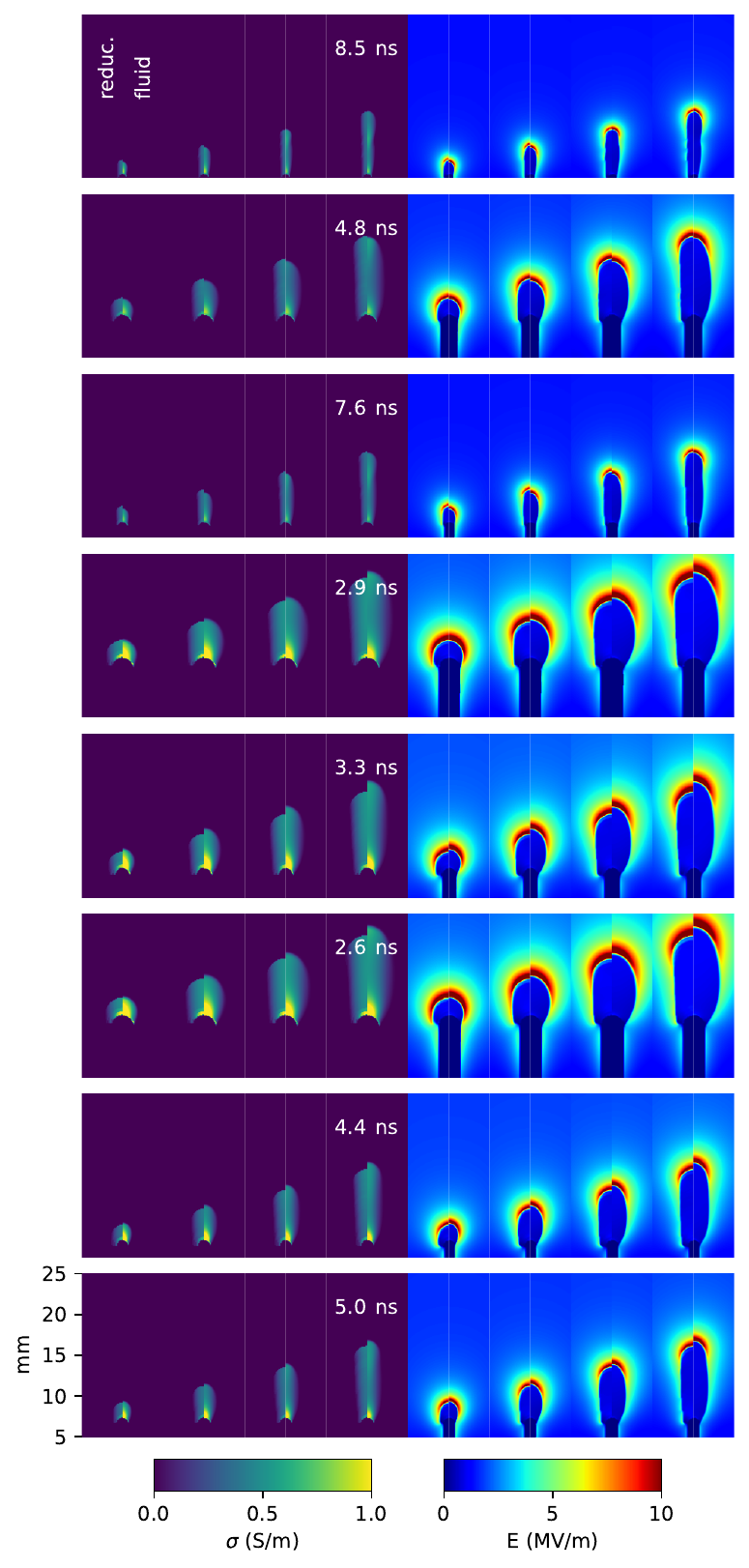}
  \caption{Comparison of reduced model against fluid simulations, with the conductivity on the left and the electric field strength on the right.
    The left-halves of the sub-plots correspond to the reduced model, and the right-halves to the fluid model.
    Each row corresponds to one case from the simulation dataset, with output shown at 5, 10, 15 and 20 times $\Delta t_\mathrm{output}$, and the time at $20 \, \Delta t_\mathrm{output}$ is shown.
    Note that the rod electrode, visible as the dark region in the electric field plots, varies in radius and in length for each case.
  }
  \label{fig:axi-comparison}
\end{figure}

\begin{figure}
  \centering
  \includegraphics[width=\linewidth]{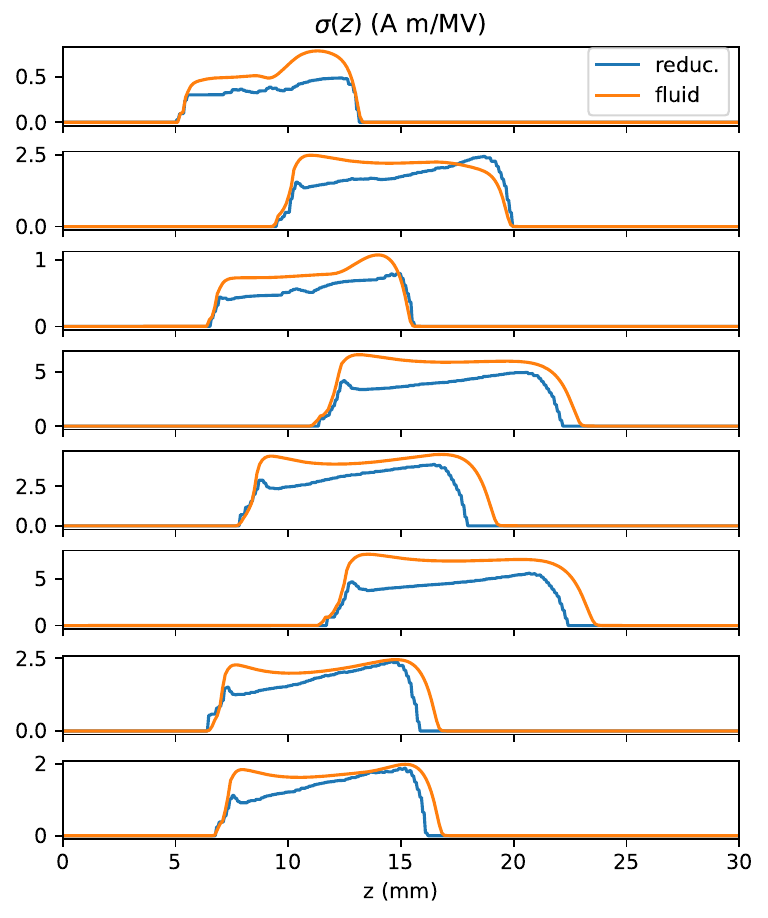}
  \caption{Comparison of line conductivity between fluid simulations and the reduced model.
    The cases shown correspond to the final time states shown in figure~\ref{fig:axi-comparison}.
  }
  \label{fig:axi-comparison-sigmaz}
\end{figure}

\begin{figure}
  \centering
  \includegraphics[width=\linewidth]{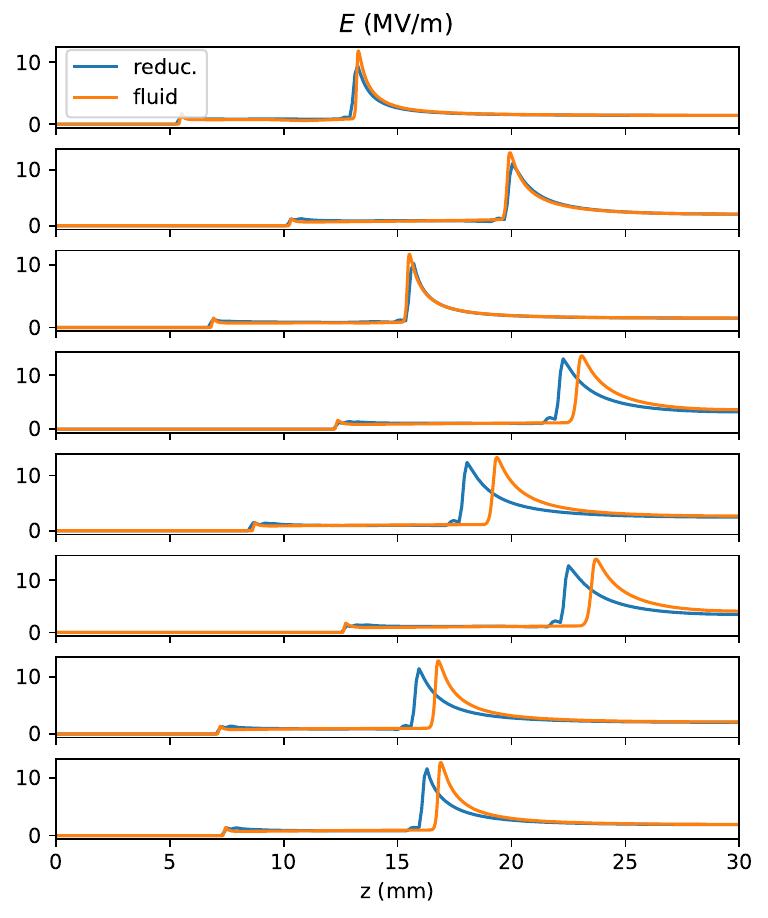}
  \caption{Comparison of on-axis electric field between fluid simulations and the reduced model.
    The cases shown correspond to the final time states shown in figure~\ref{fig:axi-comparison}.
  }
  \label{fig:axi-comparison-Ez}
\end{figure}

In figure~\ref{fig:axi-comparison} the conductivity profiles and electric field profiles resulting from both models are compared.
The cases shown are the first eight runs of the dataset.
Furthermore, figures~\ref{fig:axi-comparison-sigmaz} and \ref{fig:axi-comparison-Ez} show the corresponding line conductivities, as defined by equation~\eqref{eq:line-sigma}, and the on-axis electric field profiles.
The reduced model shows good agreement regarding streamer velocity.
The streamer radius agrees reasonably well too, and so do the electric field profiles.
However, differences can be observed as well.
For some cases, the line conductivity is almost a factor two lower near the rod electrode with the reduced model.
Furthermore, the radius profile is generally more flat in the reduced model, whereas the radius in the fluid simulations initially expands.
Such differences are not unexpected, since the reduced model cannot accurately describe streamer inception, and it is of course a quite simple model with only a single input $L_E$.

\subsection{Dependence on time step and grid resolution}
\label{sec:dependence-time-step}

\begin{figure}
  \centering
  \includegraphics[width=\linewidth]{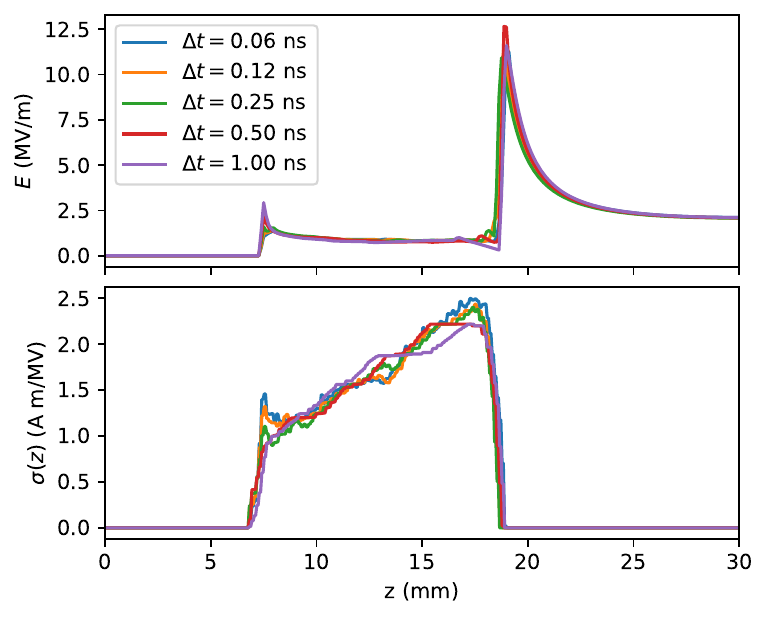}
  \caption{The dependence of the reduced model on the time step.
    Shown are on-axis electric fields and the line conductivity profiles for the bottom case of figure~\ref{fig:axi-comparison}.
    In the corresponding fluid simulation the time step was about $2 \, \textrm{ps}$.
  }
  \label{fig:axi-comparison-dt}
\end{figure}

To study how the results of the reduced model depend on the time step, we solve the bottom case shown in figure~\ref{fig:axi-comparison} using time steps ranging from $\Delta t = t_\mathrm{end}/96 = 0.0625 \, \mathrm{ns}$ up to $\Delta t = t_\mathrm{end}/6 = 1.0 \, \mathrm{ns}$, where $t_\mathrm{end} = 6.0 \, \mathrm{ns}$ (which is slightly later than the time shown in figure~\ref{fig:axi-comparison}).
The resulting line conductivity profiles and on-axis electric fields are shown in figure~\ref{fig:axi-comparison-dt}, using a uniform grid of $256^2$ cells.
For these cases, the temporal smoothing of $L_E$ was turned off by using $\beta = 1.0$ in equation \eqref{eq:L-E-beta}.

The results with the different time steps all agree rather well, showing that the reduced model is not sensitive to the time step.
There is no CFL-like restriction, and the model can handle time steps much larger than the so-called dielectric relaxation time $\tau_\mathrm{drt} = \varepsilon_0/\sigma$ due to the implicit solution of equation~\eqref{eq:phi-update-2}.
In comparison, the fluid simulations had to use a time step of about $2 \, \textrm{ps}$, which is almost three orders of magnitude smaller than the largest time step used in figure~\ref{fig:axi-comparison-dt}.

\begin{figure}
  \centering
  \includegraphics[width=\linewidth]{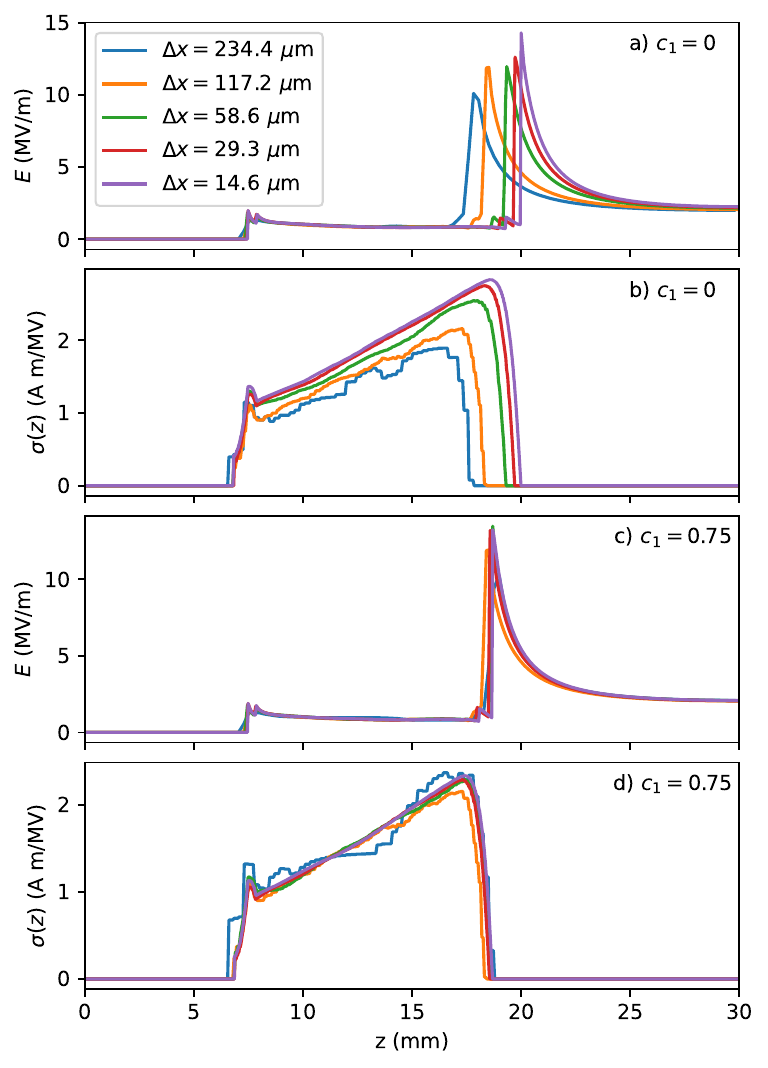}
  \caption{The dependence of the reduced model on the grid spacing.
    For a) and b) no correction was included for $L_E$, whereas equation~\eqref{eq:L-E-corrected} with $c_1 = 0.75$ was used for c) and d).
    Shown are on-axis electric fields and the line conductivity profiles for the bottom case of figure~\ref{fig:axi-comparison}.
    In the corresponding fluid simulation the finest grid spacing was $3.7 \, \mu\textrm{m}$.
  }
  \label{fig:axi-comparison-dx}
\end{figure}

In figures~\ref{fig:axi-comparison-dx}a and \ref{fig:axi-comparison-dx}b the effect of the grid resolution $\Delta x$ is illustrated.
The same test case was performed, with a time step of $0.25 \, \textrm{ns}$, but now using a grid spacing $\Delta x$ between about $15 \, \mu\textrm{m}$ ($2048^2$ uniform grid) and $234 \, \mu\textrm{m}$ ($128^2$ uniform grid).
For comparison, the fluid simulations required $\Delta x = 3.7 \, \mu\textrm{m}$ in the region around the streamer head.
The variation in grid spacing has a more significant effect on the solutions than the time step.
With a larger $\Delta x$ the streamer velocity is lower, and there is more noise in the profiles because $L_E$ cannot accurately be measured.
The maximal electric field strength is also lower with a coarser grid since the peak value is smeared out.

The main reason for the change in velocity in figures~\ref{fig:axi-comparison-dx}a and \ref{fig:axi-comparison-dx}b is that measurements of $L_E$ depend on the grid spacing $\Delta x$.
Consider for example a function that is zero for $x < x_0$ and $1- \epsilon \, x$ for $x \geq x_0$, where $x_0$ varies and $\epsilon$ is a small positive number.
If we map this function to a uniform grid with resolution $\Delta x$ \jt{by averaging its value over each grid cell} and then locate its maximum, the expected location will be $x_0 + \Delta x$, since if $x_0$ lies within a cell, the maximum will occur in the next cell.
The electric field profile near the streamer head is similarly asymmetric: towards the channel it quickly drops, whereas the decay is less steep in the streamer's propagation direction.
This will lead to a similar shift on the order of $\Delta x$, so that a corrected value $L_E^*$ can be obtained as
\begin{equation}
  \label{eq:L-E-corrected}
  L_E^* = L_E + c_1 (\Delta x - \Delta x_0),
\end{equation}
where $c_1$ is a constant and $\Delta x_0$ corresponds to the $256^2$ grid that we used to measure $L_E$ in the simulation dataset.
Using $c_1 = 1.0$ leads to a slight over-correction, since the dependence of $L_E$ on $\Delta x$ is weaker than for the `ideal' case mentioned above.
Figures~\ref{fig:axi-comparison-dx}c and \ref{fig:axi-comparison-dx}d show simulation results with $c_1 = 0.75$.
The streamer velocity is now much less affected by the grid spacing, and the resulting line conductivity profiles are highly similar.

\subsection{3D simulations with branching}
\label{sec:comp-against-exper}

We now perform 3D simulations in a different computational domain, measuring $(8 \, \textrm{cm})^3$.
A $4 \, \textrm{cm}$ long rod electrode is connected to the bottom (grounded) electrode, with a radius of $1.5 \, \textrm{mm}$.
The semi-spherical tip of the rod lies approximately at the center of the domain, so that discharges can grow in the $4 \, \textrm{cm}$ gap above it.
This geometry is qualitatively similar to the 4\,cm point-plane gaps used in~\cite{Briels_2008,Briels_2008a}.
In these experiments the tip of the needle was thinner, but the holder used to place the needle in probably had a larger radius than our rod.

For computational efficiency, the 3D simulations are performed with adaptive mesh refinement.
A minimum grid spacing of $104 \, \mu\textrm{m}$ is used where $E \geq 30 \, \textrm{kV/cm}$.
Where $E < 20 \, \textrm{kV/cm}$ the mesh was derefined \jt{by a single level, so that the grid spacing is twice as large.
More derefinement is possible, but then thin channels are not well resolved for visualization purposes.
In contrast to conventional streamer simulations, there is no CFL condition, so the time step $\Delta t = 0.5 \, \textrm{ns}$ is kept fixed regardless of the mesh spacing.}

At $t = 0$, streamers with an initial radius $R_{\sigma,0} = 0.5 \, R_{\sigma}(L_E)$ are placed at the tip of the needle electrode as discussed in section~\ref{sec:comp-against-axisymm}.
However, we now start with multiple initial streamers, since in experiments it is also common for multiple streamers to form near a needle electrode, see e.g.\ \cite{Briels_2008a}.
The initial streamer velocities are determined in the same way as after a branching event, see section~\ref{sec:branching-model}.
A random angle $\gamma$ and a random axis are used for each initial streamer, with the `parent' direction assumed to be parallel to the rod's axis.
\jt{In future work, the model could be extended with a stochastic criterion for streamer inception in regions with high electric field strength, instead of the manual placement of the initial streamers done here.}

\begin{figure}
  \centering
  \includegraphics[width=\linewidth]{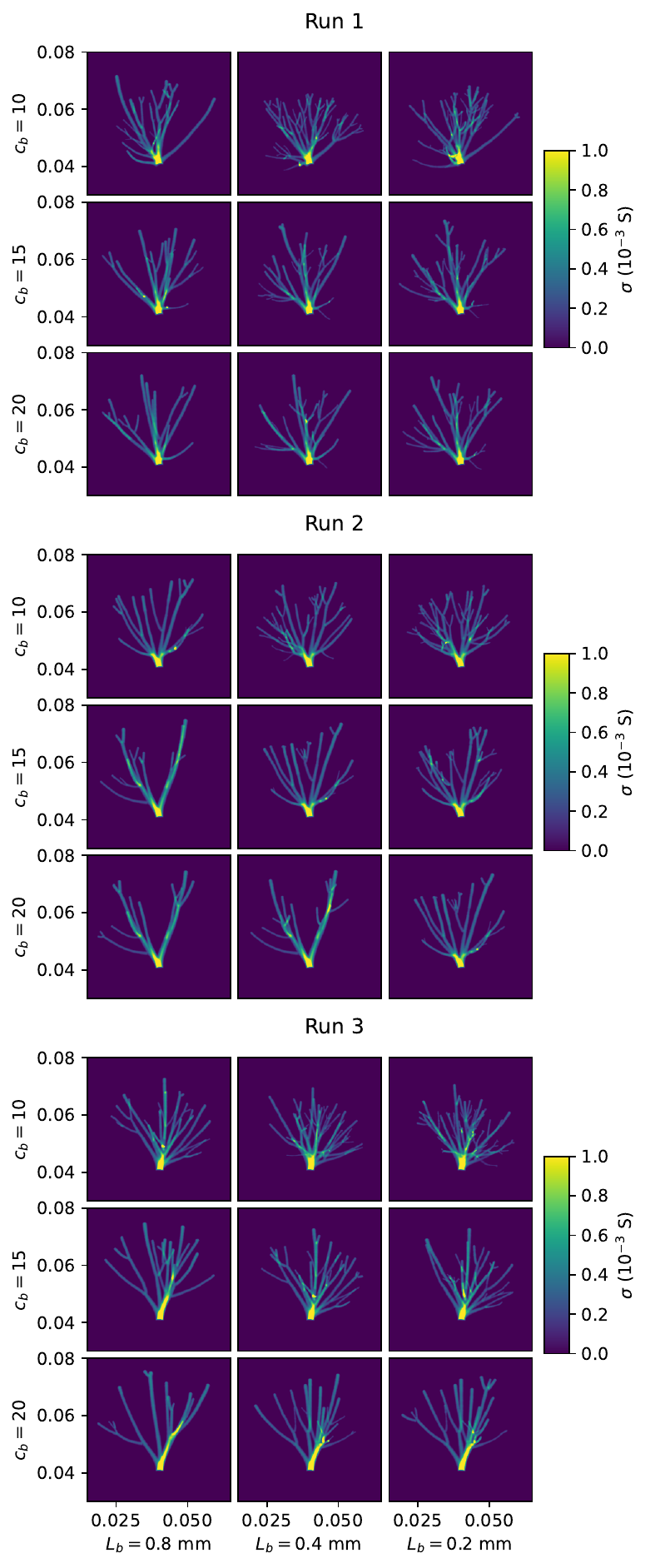}
  \caption{3D simulations of branching positive streamers in air with the reduced model. A needle at $40 \, \textrm{kV}$ was placed $4 \, \textrm{cm}$ below the grounded top plate. Shown is the conductivity $\sigma$ at $t = 35 \, \textrm{ns}$ integrated along the line-of-sight. The branching parameters $c_b$ and $L_b$ are varied. For each case, three runs are shown, which differ only in the sequence of pseudo-random numbers used.}
  \label{fig:3d-branching}
\end{figure}

In figure~\ref{fig:3d-branching} simulation results with an applied voltage of $40 \, \textrm{kV}$ and five initial streamers are shown.
The branching parameters $c_b$ and $L_b$ are varied in this figure, with $c_b$ between 10 and 20 and $L_b$ between $0.2 \, \textrm{mm}$ and $0.8 \, \textrm{mm}$.
The directions of the initial streamers as well as branching processes depend on pseudorandom numbers.
For each case we have performed three runs, with three different sequences of random numbers.

With larger values of $c_b$ and $L_b$ there is less branching.
A difference is that increasing $L_b$ reduces the branching (and thus formation) of thin channels. If $L_b$ were zero, the number of thin branches would grow rapidly with discharge size, since the branching length $L_\mathrm{branch}$ would then be proportional to the channel radius, which tends to decrease as more branches form.

We can qualitatively compare our results with the experiments of~\cite{Briels_2008} with a $+40 \, \textrm{kV}$ voltage in a $4 \, \textrm{cm}$ gap.
In the simulations the voltage is applied instantaneously, so we compare against the results with relatively fast voltage rise times of $25$ to $30 \, \textrm{ns}$ in~\cite{Briels_2008}.
The velocity of the fastest simulated streamers is about $1.1 \, \textrm{mm/ns}$, whereas the fastest velocities in the experiments were about $0.8 \pm 0.2 \, \textrm{mm/ns}$.
The main reason for this discrepancy is probably the voltage rise time used in the experiments, which is comparable to the time it takes streamers to cross the gap.
In~\cite{Briels_2008} it was also shown that with a shorter rise time, fewer but thicker streamers form, which propagate faster.
Due to the dependence of the discharge morphology on the rise time in the experiments, it is difficult to determine which values of $c_b$ and $L_b$ best agree with the experiments.
However, there appears to be too little branching in the simulations when $L_b = 0.8 \, \textrm{mm}$ or when $c_b = 20$.
With $c_b = 10$ and $L_b = 0.2 \, \mathrm{mm}$ there is perhaps a bit too much branching, so we think that for example $c_b = 15$ and $L_b = 0.2 \, \mathrm{mm}$ could be reasonable.

Qualitatively, the reduced model simulations reproduce several phenomena seen in the experiments.
Thin branches that are overtaken by nearby channels tend to stagnate, which limits the number of such thin branches.
Furthermore, due to electrostatic repulsion, some channels that form near the electrode tip propagate almost horizontally. These channels are slower than the vertically propagating channels.
Also note that the region near the electrode tip clearly has the highest conductivity.
A qualitative difference is that in experiments with a fast rise time, a so-called `inception cloud' can form around the electrode tip, see e.g.\ \cite{Chen_2015}, which does not happen in the reduced model.

\section{Discussion}
\label{sec:discussion}

\subsection{Line conductivity versus line electron density}
\label{sec:line-conductivity}

In the framework described in section~\ref{sec:framework-description} and in the reduced model described in section~\ref{sec:model-description}, we use a line conductivity $\sigma_\mathrm{line}$.
The conductivity depends on both the electron density and the electron mobility, see equation~\eqref{eq:line-sigma}.
The dependence of the mobility on the electric field is known, as shown in figure~\ref{fig:alpha-mu}.
From a physical point of view, it could therefore make sense to instead use the `line electron density', obtained by integrating the electron density over a channels' cross-section.
The line conductivity could then be obtained by multiplying with a field-dependent mobility.
However, the main drawback of such an approach is that we have to predict the future conductivity in equation~\eqref{eq:phi-update-2} without knowing the electric field, which would render this equation implicit.

\subsection{Pros and cons of using a numerical mesh}
\label{sec:mesh-pros-cons}

An important difference between our approach and the tree model presented in~\cite{Luque_2014} is that we evolve the conductivity of the discharge on a numerical mesh, whereas channels are approximated by a series of point or ring charges in the tree model, between which pair-wise electrostatic interactions are computed.
An advantage of using a numerical mesh is that a standard electrostatic field solver can be used, so that existing methods for including electrodes, dielectrics and mesh refinement can be re-used.
Since we resolve the streamer radius on the numerical mesh, there is also no need for the custom interaction kernels of the tree model~\cite{Luque_2014,Luque_2017a}.
A drawback of the mesh-based approach is that a relatively fine mesh is still required to simulate thin streamers.
However, in terms of computational cost our model does not seem slower than the tree model, since 3D simulations of similar complexity as the ones shown in section~\ref{sec:comp-against-exper} were reported to take a few hours on a modern desktop computer with the tree model~\cite{Luque_2014}.
Another benefit of a mesh-based approach is that it is relatively easy to couple simulations to gas dynamics, which could be relevant to study the streamer-to-leader transition.

\subsection{Computational cost of 3D simulations}
\label{sec:computational-cost}

The computational cost of the 3D simulations presented in section \ref{sec:comp-against-exper} was about 4-8 minutes per case, with the simulations running up $t = 40 \, \textrm{ns}$ on a desktop computer with an AMD 2700X 8-core CPU.
The majority of this CPU time was spent on solving the variable-coefficient Poisson's equation~\eqref{eq:phi-update-2}.
Writing output to a hard disk at every time step also took quite some time, since each run produced about 10-20 GB of data.
Due to the use of adaptive mesh refinement, the simulations were more expensive when there were more branches, covering a larger volume.
Up to about $10^7$ grid cells were used.
We remark that the simulations could have been sped up by allowing for more mesh derefinement.

\subsection{Conclusions and outlook}
\label{sec:conclusion}

We have presented a computational framework for the simulation of filamentary electric discharges, in which the channels are represented as conducting cylindrical segments.
The input for this framework is a model that predicts the position, radius and line conductivity of each channel at the next time step.
This information is used to update the electric conductivity on a numerical mesh, from which the electric potential at the next time step is computed by solving a variable-coefficient Poisson equation.
An existing field solver~\cite{Teunissen_2018} is used for this purpose, which supports adaptive mesh refinement and parallelization.
To allow for easy experimentation with different types of models, the computationally expensive components of the framework are written in Fortran but compiled into a Python module.

To demonstrate how the framework can be used, we have generated a dataset of 1000 axisymmetric positive streamer simulations in air at 1\, bar, by varying the applied voltage and the electrode geometry.
The simulations were performed with a drift-diffusion fluid model and the local field approximation, using applied voltages between $36 \, \textrm{kV}$ and $60 \, \textrm{kV}$ in a $30 \, \textrm{mm}$ gap.
The needle electrode length was varied between $4.5 \, \textrm{mm}$ and $12 \, \textrm{mm}$.
Based on the simulation data, simple fit expressions were presented for the streamer radius, velocity and line conductivity.
These fits depend on a single input feature, which measured the size of the high-field region ahead of the streamer.

Based on the computational framework and the fitted expressions, a reduced model for positive streamers in air was presented.
In this model, streamers are assumed to propagate parallel to the electric field, and branching is included a stochastic process described by two parameters.
The reduced model was compared against axisymmetric simulations from the dataset, and it generally showed good agreement despite its simplicity.
We demonstrated that the spatial and temporal step sizes can be several orders of magnitude larger in the reduced model than in the classical fluid model, and that the reduced model is not sensitive to these step sizes.

Finally, we have presented 3D simulations with the reduced model, using a 4\,cm needle-plate gap with an applied voltage of 40\,kV, similar to the geometry used in some of the experiments presented in~\cite{Briels_2008}.
In the simulations the streamer velocity was about 30\% higher than in the experiments, which we mainly attributed to the voltage rise time in the experiments.
The effect of the branching parameters was studied, and similar discharge morphologies were observed as in the experiments.
However, it should be mentioned that the experimental results were quite sensitive to the voltage rise time, so we could not accurately determine values for the branching parameters.
The cost of the 3D simulations was rather modest, with individual runs taking 4-8 minutes on a desktop computer, depending on the complexity of the discharge.

The parameters of the model presented here could be \rev{improved} by comparing against experiments, in particular regarding streamer branching.
Perhaps similar models can also be developed for gases other than air, in which branching is typically more frequent.
In future work, it would also be interesting to experiment with different discharge models, for example a physics-based reduced model~\cite{Bouwman_2024} or a more sophisticated machine-learning model.
Negative streamers could also be considered in the future.
Another option would be to study the streamer-to-leader transition with the reduced model, which could then be coupled to equations for gas dynamics.

\section*{Acknowledgements}

A.M.R.~was supported by a Ram\'{o}n Areces Foundation grant No.~BEVP34A6840.
The reduced discharge model described in this paper is available at \url{https://github.com/jannisteunissen/cocydimo} and the simulation dataset is available at \url{https://doi.org/10.5281/zenodo.15296406}

\appendix

\section{Comparison of different streamer radius definitions}
\label{sec:comp-radius}

\begin{figure}
  \centering
  \includegraphics[width=\linewidth]{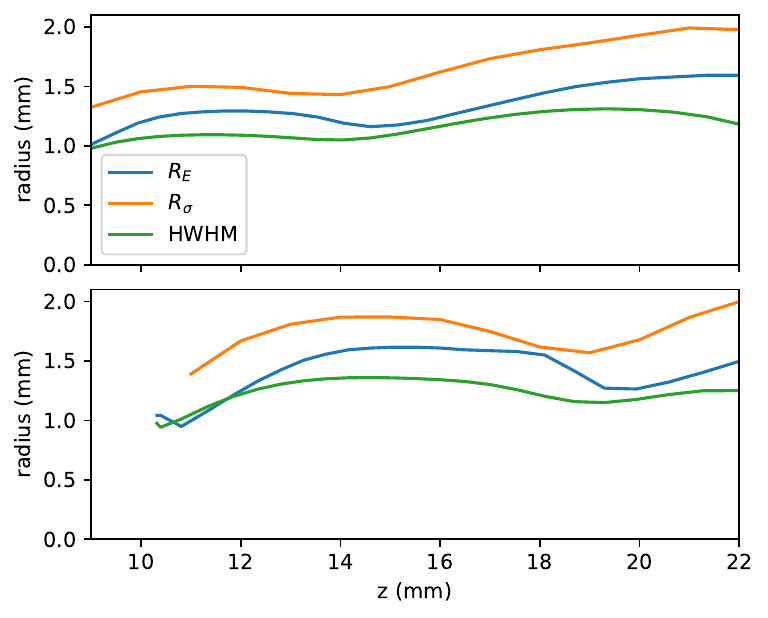}
  \caption{Comparison of different definitions of the streamer radius (see text) versus the $z$-coordinate of the streamer head, for two of the cases from the dataset described in section~\ref{sec:description-data-set}.}
  \label{fig:radius-comparison}
\end{figure}

\begin{figure}
  \centering
  \includegraphics[width=\linewidth]{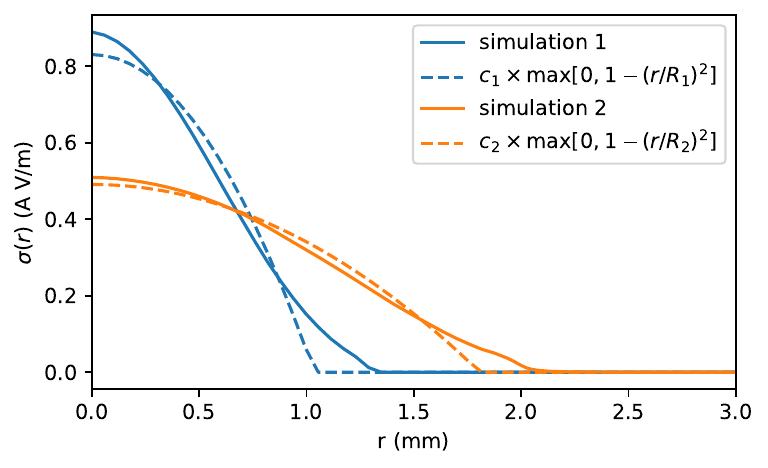}
  \caption{Two examples of radial conductivity profiles from simulations together with fitted curves assuming a radial profile as given by equation~\eqref{eq:fr-function}.}
  \label{fig:radial-profile}
\end{figure}

There are quite a few different definitions of the streamer radius, see e.g.\ \cite{Pancheshnyi_2005,Guo_2022b}.
We here consider three definitions of the radius:
\begin{itemize}
  \item The electrodynamic radius $R_E$, defined as the radial coordinate at which the radial electric field $E_r$ has a maximum.
  \item The radius $R_\sigma$ of the streamer's radial conductivity profile, found by fitting the radial conductivity profile with a function of the form
  \begin{equation}
    \label{eq:fit-func-radial}
    f(r) = c_0 \, \max\left(0, 1 - (r/R_\sigma)^2\right),
  \end{equation}
  similar to equation \eqref{eq:fr-function}.
  \item The optical radius, defined as the ``Half Width at Half Maximum'' (HWHM) of the time-integrated optical emission, after applying a Forward Abel transform on the axisymmetric simulation data.
\end{itemize}
Figure \ref{fig:radius-comparison} shows a comparison of these definitions for two of the simulations in our dataset.
It can be seen that $R_\sigma$ is typically about 10\% to 30\% larger than $R_E$, and that $R_\sigma$ is typically about 30\% to 40\% larger than HWHM.
Based on these observations, we use the rather rough approximation $R_\sigma = 1.2 \, R_E$ in this paper.
Finally, figure~\ref{fig:radial-profile} shows two examples of radial conductivity profiles from our simulation dataset together with fitted curves according to equation~\eqref{eq:fit-func-radial}.


\bibliography{zotero_jannis,other}
\bibliographystyle{unsrt}

\end{document}